\documentstyle[aps,twocolumn]{revtex}

\begin{document}
\title{Magnetic dilution in the geometrically frustrated SrCr$_{9p}$Ga$_{12-9p}$O$%
_{19}$ and the role of local dynamics: a $\mu $SR study}
\author{A.~Keren$^{1}$, Y.~J.~Uemura$^{2}$, G. Luke$^{3}$, P.~Mendels$^{4}$,
M.~Mekata$^{5}$ and T.~Asano$^{6}$}
\address{$^{1}$Department of Physics, Technion - Israel Institute of Technology,\\
Haifa 32000, Israel.\\
$^{2}$Physics Department, Columbia University, New York City, New York 10027.%
\\
$^{3}$Department of Physics \& Astronomy, McMaster University, 1280 Main St.%
\\
West Hamilton, Ontario L8P 4M1, Canada\\
$^{4}$Laboratoire de Physique des Solides, URA2 CNRS, Universit\'{e} Paris\\
Sud, 91405 Orsay, France\\
$^{5}$Department of Applied Physics, Fukui University, Fukui 910, Japan.\\
$^{6}$Department of Physics, Kyushu University, Fukuoka 812-8581, Japan}
\date{\today}
\maketitle

\begin{abstract}
We investigate the spin dynamics of SrCr$_{9p}$Ga$_{12-9p}$O$_{19}$ for $p$
below and above the percolation threshold $p_{c}$ using muon spin
relaxation. Our major findings are: (I) At $T\rightarrow 0$ the relaxation
rate is $T$ independent and $\propto p^{3}$, (II) the slowing down of spin
fluctuation is activated with an energy $U$ which is also a linear function
of $p^{3}$ and $\lim_{p\rightarrow 0}U=8$~K; this energy scale could stem
only from a single ion anisotropy, and (III) the $p$ dependence of the
dynamical properties is identical below and above $p_{c}$, indicating that
they are controlled by local excitation.
\end{abstract}

The SrCr$_{9p}$Ga$_{12-9p}$O$_{19}$ [SCGO($p$)] compound has been intensely
studied experimentally in recent years, as it is believed to be a model
compound for a ``super-degenerate'' antiferromagnet \cite
{BroholmPRL90,MartinezPRB92,LeePRL96,KerenHypInt94,UemuraPRL94,Mendels,RamirezPRB92,SchifferPRL96,RamirezPRL90}%
. By ``super-degenerate'' we mean that the classical ground state energy is
invariant under a local rotation of a small number of spins, a symmetry
which leads to local zero energy excitations in addition to the more common
collective ones. One of the open questions in this area of research is which
type of excitations will dominate in low temperatures. \ To date, this
question was addressed only theoretically in two of the most famous
super-degenerate magnets, the kagom\'{e} \cite{KerenPRL94} and the
pyrochlore \cite{MoessnerPRB98}, where it was found that the dynamical
properties are mostly controlled by the local excitations. Here we examine
this question experimentally. We do so by measuring muon spin relaxation ($%
\mu $SR) rates as a function of temperature $T$, magnetic field $H$, and,
most importantly, magnetic ion concentration $p$ above and below the
percolation threshold $p_{c}$. Our major finding is that the dynamical
properties of the system are impartial to $p_{c}$, suggesting that they
could not emerge from a collective phenomenon.

Unexpectedly, our measurements also lead us to another finding regarding the
spin Hamiltonian in SCGO. For long it was suspected that this Hamiltonian
must contain terms other than the Heisenberg one. This is because SCGO shows
spin glass like effect in susceptibility experiments, such as a cusp at a
temperature $T_{f}\simeq 4p$~K \cite{MartinezPRB92}, and a hysteresis
between the field cooled (FC) and zero field cooled (ZFC) measurements.
Neither of these could be understood in terms of Heisenberg spins on
kagom\'{e} or pyrochlore lattices \cite{ShenderFO94}. Our data indicate the
presence of a single ion anisotropy with an energy scale of $\sim 8$~K which
might be the origin of the susceptibility effects.

Our samples were prepared by solid state reaction at 1350C from the
stoichiometric mixtures of Cr$_{2}$O$_{3}$, Ga$_{2}$O$_{3}$ and SrCO$_{3}$.
X-ray examination revealed the absence of foreign phase in the prepared
samples and a slight smooth lattice expansion as $p$ is decreased. Our
intention was to have $p$ values both above and below the $p_{c}$ of the
lattice. However, there is a controversy whether SCGO represents a kagom\'{e}
lattice or a pyrochlore-slab \cite{LeePRL96}. The $p_{c}$ for the kagom\'{e}
lattice is $0.6527$ and for a pyrochlore-slab is not known. Nevertheless,
the Curie-Weiss temperature $\Theta _{\text{cw}}$ in SCGO($p$) is a linear
function of $p$ with a slope which is different below and above $p=0.61(5)$ 
\cite{MartinezPRB92}, in agreement with $p_{c}$ for the kagom\'{e} lattice.
Therefore we assume that $p_{c}=0.6527$ and prepare samples with $p$ in the
range $0.39$ to $0.89$. The value of $p$'s in our samples is determined from
their Curie constant. This method was found to be in agreement with the
stoichiometric ratio in the sample preparation, the Curie-Weiss temperature,
and $T_{f}$ (when measurable) \cite{MartinezPRB92}.

Our $\mu $SR experiments were done in both TRIUMF and ISIS. In these
experiments one follows the time evolution of the spin polarization $%
P_{z}(t) $ of a muon implanted in a sample, through the asymmetry $%
A(t)\propto P_{z}(t)$ in the positron emission of the muon decay. In
addition, an external field $H$ is applied along the initial muon spin
(longitudinal) direction which defines the ${\bf z}$ axis. $A(t)$ for three
different samples at base temperature and $H=100$~G is shown by the symbols
in Fig.~\ref{AsyVsT} where time is presented on a log scale. The small field
of $100$~G could be considered as zero field; it is applied in order to
decouple the nuclear spin contribution to the muon spin relaxation. As can
be seen from the figure, there is a strong variation in the time scale of
relaxation between the different samples. In addition, the asymmetry for all
samples has a ``flat'' beginning similar to Gaussian. In the inset of Fig.~%
\ref{AsyVsT} we depict the asymmetry for the $p=0.89$ at high temperature ($%
5.5$~K) where $A$ is presented on a log scale. Clearly at high temperatures
the relaxation is closer to exponential. Therefore, the asymmetry waveform
is temperature dependent, as was first found in the $p=0.89$ sample by
Uemura {\it et al.} \cite{UemuraPRL94}, but depends weakly on the most
important variable in this paper, namely, $p$.

There are two ways to obtain a relaxation rate in a situation where the
waveform is changing: (I) using the $1/e$ criteria where we define the time $%
T_{1}$ by $A(T_{1})=A(0)/e$, or (II) fitting all data sets to a stretched
exponential 
\begin{equation}
A(t)=A_{0}\exp \left( -(\lambda t)^{\beta }\right) .  \label{Asymmetry}
\end{equation}
Representative fits are depicted in Fig.~\ref{AsyVsT} by the solid lines.
However, in our case, where the stretched exponential fits the data well for
more than $1/e$ of the initial asymmetry, as demonstrated by the arrow in
Fig.~\ref{AsyVsT}, one finds that $\lambda =1/T_{1}$. Therefore, both
methods lead to the same relaxation rate although in (II) both $\lambda $
and $\beta $ are functions of $T$, $H$, and $p$ whereas in (I) these
parameters impact only $T_{1}$. We therefore continue the discussion using
the stretched exponential fit approach which is more informative and
accurate. Finally, in the samples with large $p$, Eq.~\ref{Asymmetry}
accounts only for early times where the asymmetry drops by $\sim $80\%. At
later times a second component with low relaxation rate dominates. However,
this second component is not observable in the small $p$ samples due to
experimental limitations imposed by the muon life time (2.2 $\mu $sec).
Therefore, we concentrate here on the early time behavior of $A(t)$.

In Fig.~\ref{LamVsTmZF} we show $\lambda $ in $H=100$~G as a function of
temperature for various values of $p$. All samples show critical slowing
down starting at $T=20-5$~K, followed by a saturation of the muon relaxation
rate, in agreement with earlier $\mu $SR work \cite
{KerenHypInt94,UemuraPRL94} and more recent NMR measurements \cite{Mendels}
on the $p=0.89$ sample. In Fig.~\ref{BetVsTmAndH}a we depict $\beta $ versus 
$T$ for the different $p$'s. As demonstrated before, the waveform in all our
samples is exponential ($\beta \rightarrow 1$) at high $T$ but tends to be
Gaussian ($\beta \rightarrow 2$) at low $T$.

There are two possible mechanisms which can be responsible for the loss of
the muon polarization: static field distribution, or dynamic field
fluctuations. It is possible to distinguish between these two by measuring
the field dependence of the muon spin relaxation rate. In Fig. \ref{LamVsH}
we show $\lambda (H)$ on a semi-log scale for all our samples, and in Fig.~%
\ref{BetVsTmAndH}b the field dependence of $\beta $. The field dependence of 
$\lambda $ allows us to rule out the possibility of relaxation due to static
field distribution as the following argument shows. In the $p=0.89$ sample,
when $H=100$~G, and $T\rightarrow 0$, the value of $\lambda $ is $10$~$\mu $%
sec$^{-1}$. If this relaxation would have been due to static field
distribution it would have implied an internal magnetic field $[B]$\ at the
muon site in the order of $100$~G (using $\lambda \sim \gamma _{\mu }[B]$
where $\gamma _{\mu }=85.16$ MHz/kG). The vector sum of this internal field
and an external longitudinal field of, say, $2$~kG, would have been nearly
parallel to the initial muon spin direction. Therefore, if the internal
fields were static, we would expect a complete quenching of the relaxation
rate in $2$~kG and above, in contrast to observation. This line of argument
applies to all other samples as well. Thus, the decay of $P_{z}(t)$ is not
due to static random fields, and seems to be due to dynamic field
fluctuations.

On the other hand, the waveform is a Gaussian with a $\lambda $ that has a
very weak field dependence for $H$ up to $2$~kG; a field that obeys $\gamma
_{\mu }H/\lambda \gg 1$. This stands in strong contrast to all theories
known to us of relaxation from dynamical fluctuations. These theories yield
exponential relaxation when there is weak field dependence for $\gamma _{\mu
}H/\lambda \gg 1$. A dynamical Gaussian waveform is also very unusual
experimentally, and is one of the on going puzzles of $\mu $SR in SCGO \cite
{UemuraPRL94}. Nevertheless, we interpret our data using a dynamical model,
since the argument regarding the vector sum of internal and external fields
seems more fundamental than the exact waveform.

In dynamical models $\lambda (H)$ is proportional to the Fourier transform
of the local field dynamical auto correlation function $\left\langle
B_{i}(t)B_{j}(0)\right\rangle $ $(i=x,$ $y,$ $z)$ at the frequency $\omega
=\gamma _{\mu }H$. We find that for all samples $\lambda (H)\propto \exp
(-H/H_{0})$ where $H_{0}=1.16$~T, as demonstrated by the solid line in Fig.~%
\ref{LamVsH}. This fact indicates that the spectral density is not modified
by magnetic dilution, apart from an over all factor, and that it is
impartial to percolation.

When no external field is applied the internal field fluctuation rate $\nu $
is related to the muon relaxation rate by 
\begin{equation}
\nu \propto \left\langle B^{2}\right\rangle \lambda ^{-1}
\label{FluctuationRate}
\end{equation}
where $\left\langle B^{2}\right\rangle $ is the rms of the instantaneous
field distribution at the muon site. We find that $\lambda ^{-1}(T)$ could
be well fitted to 
\begin{equation}
\lambda ^{-1}(p,T)=Q(P)+C\exp \left[ -U(p)/T\right]  \label{InvRlxRate}
\end{equation}
where $C=150(10)$ $\mu $sec is a global parameter. The fit results are given
in Fig.~\ref{LamVsTmZF} by the solid lines. This fit suggests that the
internal field fluctuation rate is controlled by two dynamical processes: a
temperature independent quantum one, and an activated one. A similar
behavior was observed in the high spin molecul system CrNi$_{6}$ where high $%
T$ dynamics is due to over-the-barrier motion, and low $T$ dynamics is due
to quantum fluctuations \cite{KerenRCom99}. Surprisingly, we find that $%
Q^{-1}(p)$ and $U(p)$ are linear functions of $(p/p_{c})^{3}$ both below and
above $p_{c}$ as demonstrated in Fig.~\ref{QpAndUpVsP}. This result,
together with the field dependence of $\lambda $, leads us to our first main
conclusion, namely, that the fluctuations are impartial to whether the
lattice percolates and supports collective excitations or not. Therefore,
the excitations are of local nature.

It is interesting to compare this finding with other relevant experiments.
Heat capacity measurements were performed by Ramirez {\it et al.} \cite
{RamirezPRB92}. They found that $C(T)\sim T^{2}$ even when the percolation
threshold for the kagom\'{e} lattice was crossed. They pointed out that this
temperature dependence could result from collective antiferromagnetic
excitations in two dimensions of the acoustic type, namely, $\omega \propto
k $. A similar indication was made by Broholm {\it et al.} \cite
{BroholmPRL90} based on density of states $\rho (\omega )\propto \omega $
found in neutron scattering. However, no dispersion relation of the acoustic
type was ever found. Our finding of local excitations indicates that at low $%
T$ the important $\omega $'s are $k$ independent.

An extrapolation of $Q^{-1}(p)$ in Fig.~\ref{QpAndUpVsP} to $p=0$ suggests
that $Q$ diverges upon dilution. In fact for $p\leq 0.05$ we expect $Q(p)\gg
C$ and the muon relaxation rate $\lambda $ should be $T$ independent. In
other words, SCGO($p\leq 0.05$) should behave as if its spins are isolated.
A similar extrapolation of $U(p)$ to low $p$ gives $8$~K. This leads to our
second major finding, namely, that the activated part of the dynamics does
not emerge only from coupling between neighboring spins, but also from on
site (single ion) interactions. The energy scale of this interaction is $8$%
~K. When comparing this result with other experiments we find that it agrees
with the energy scale of the anisotropy found by Schiffer {\it et al.} in
their susceptibility measurements on SCGO single crystal \cite{SchifferPRL96}%
, but disagrees with Ohta{\it \ et al.} who found a single ion energy two
orders of magnitude smaller using EPR \cite{OhtaPSJ96}.

The conclusions drawn up to now are based on gross features in the data. Now
we would like to speculate on how the $p$ dependence of the muon relaxation
rate $\lambda (p)$ is shared between the instantaneous field distribution $%
\left\langle B^{2}\right\rangle (p)$ and the fluctuation rate $\nu (p)$
which determines it using Eq.~\ref{FluctuationRate}. First we calibrate $%
\left\langle B^{2}\right\rangle (p)$ from the high temperature data where $%
\lambda $ shows a weak $p$ dependence (See Fig.~\ref{LamVsTmZF}). We assume
that $\lambda \propto p^{\epsilon }$ for large $T$, where $\epsilon $ is a
small number. In addition, it is natural to expect $\nu \propto Jp^{1/2}$
(where $J$ is a coupling constant) in the high temperature range. Therefore,
our calibration leads to $\left\langle B^{2}\right\rangle (p)\propto
p^{1/2+\epsilon }$, an expression which is only slightly different from the
dilute limit where $\left\langle B^{2}\right\rangle (p)\propto p$ ($\epsilon
=1/2$). In SCGO the relation $\left\langle B^{2}\right\rangle (p)\propto
p^{1/2+\epsilon }$ should hold at all temperatures since no static moment
develops. Therefore, at base temperature, where $\lambda \propto p^{3}$ we
expect $\nu \propto p^{\epsilon -2.5}$, and since we can overrule small
values of $\epsilon $ there is a reasonable chance that $\nu \propto p^{-2}$
.

In summary the dynamical spin fluctuations in SCGO are controlled by both
quantum and activated dynamical processes. The activation energy is a linear
function of $p^{3}$, and indicates the presence of single ion anisotropy
with an energy scale of $8$~K. This anisotropy might be responsible for the
spin glass like behavior of SCGO. The quantum fluctuation time ($1/\nu $) is
most likely proportional to $p^{2}$. Both of the dynamical properties are
impartial to the percolation threshold and indicate the local nature of the
fluctuations.

We are grateful for the technical support of J.~Lord from ISIS and C.
Ballard from TRIUMF. A. Keren, Y. J. Uemura, and G. Luke would like to thank
the Israel - U. S. Binational Science Foundation and Y. J. Uemura and G.
Luke appreciate the NSF-DMR-98-02000 (Columbia), and NEDO International
Joint Research Grants for supporting their research.



\begin{figure}[tbp]
\caption{The asymmetry in the emission of muon decay positrons $\propto
P_{z}(t)$ as a function of time, for three different samples at base
temperature, is presented on a semi-log scale by the symbols. The solid
lines are fits to Eq.~\ref{Asymmetry}. The inset shows the asymmetry at high
temperature, again on a semi-log scale. The solid line represents
exponential decay.}
\label{AsyVsT}
\end{figure}

\begin{figure}[tbp]
\caption{Muon spin relaxation rates $\protect\lambda $ as obtained from fits
of raw data to Eq.~\ref{Asymmetry} for various values of $p$. The solid
lines are fits to Eq.~\ref{InvRlxRate}.}
\label{LamVsTmZF}
\end{figure}

\begin{figure}[tbp]
\caption{The exponent $\protect\beta $ as obtained from a fit of the raw
data to Eq.~\ref{Asymmetry} for various values of $p$, (a) as a function of $%
T$ at $H=100$~G, and (b) as a function of $H$ at base temperature.}
\label{BetVsTmAndH}
\end{figure}

\begin{figure}[tbp]
\caption{The field dependence of the muon relaxation rate, plotted on a
semi-log scale, for various values of $p$. The solid lines are parallel and
are described in the text.}
\label{LamVsH}
\end{figure}

\begin{figure}[tbp]
\caption{The parameters $Q^{-1}$ and $U$ of Eq.~\ref{InvRlxRate} plotted vs. 
$(p/p_c)^3$.}
\label{QpAndUpVsP}
\end{figure}

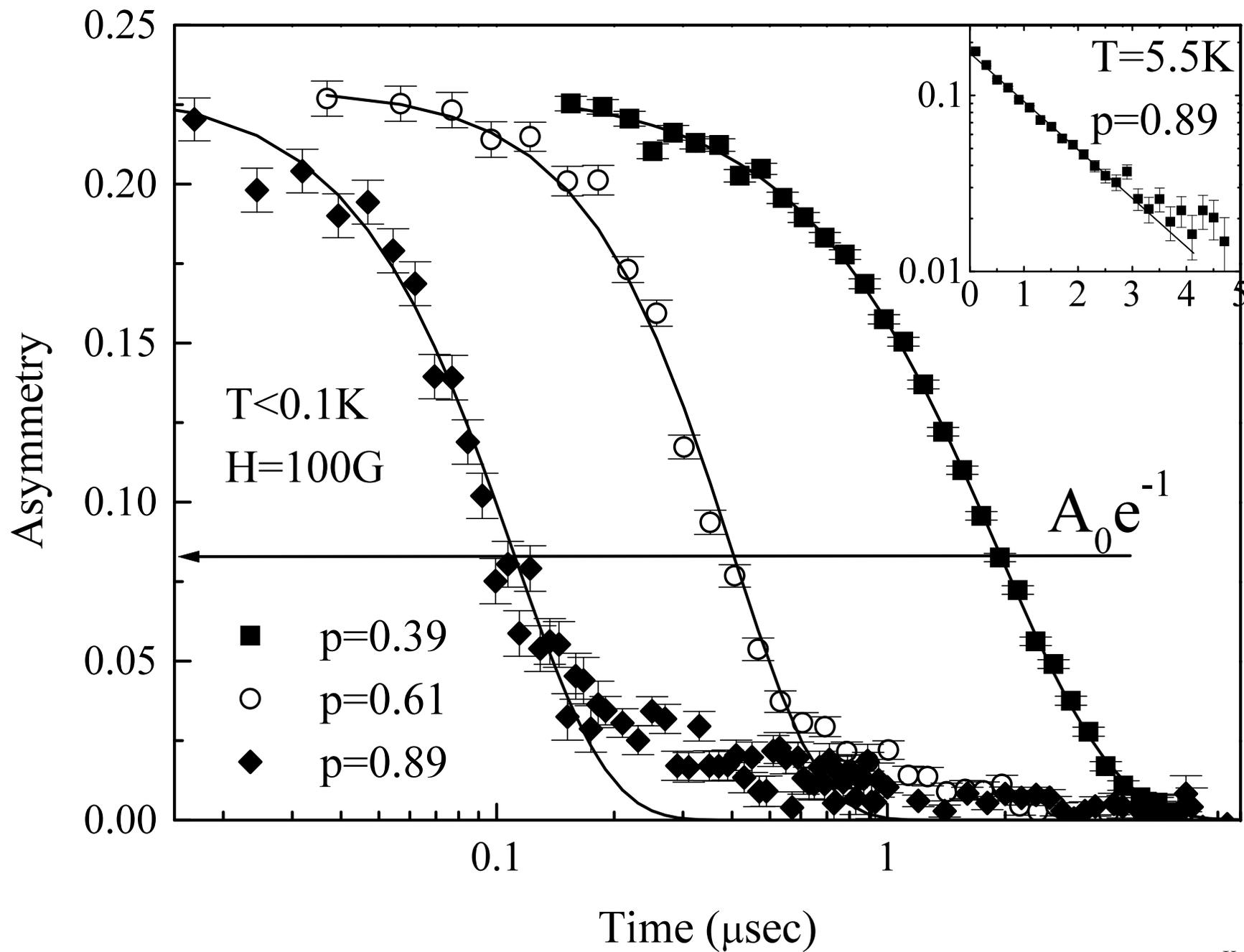
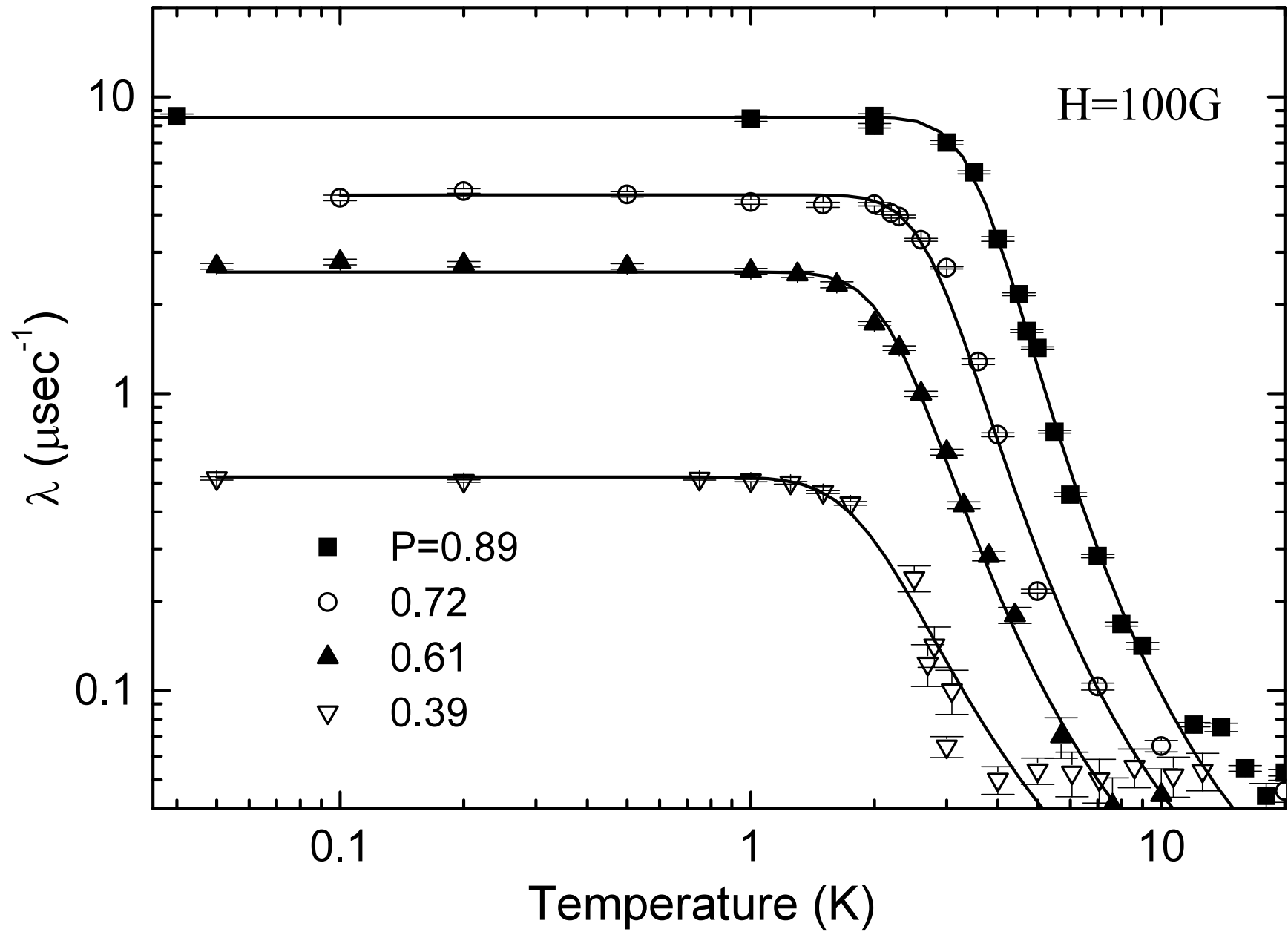
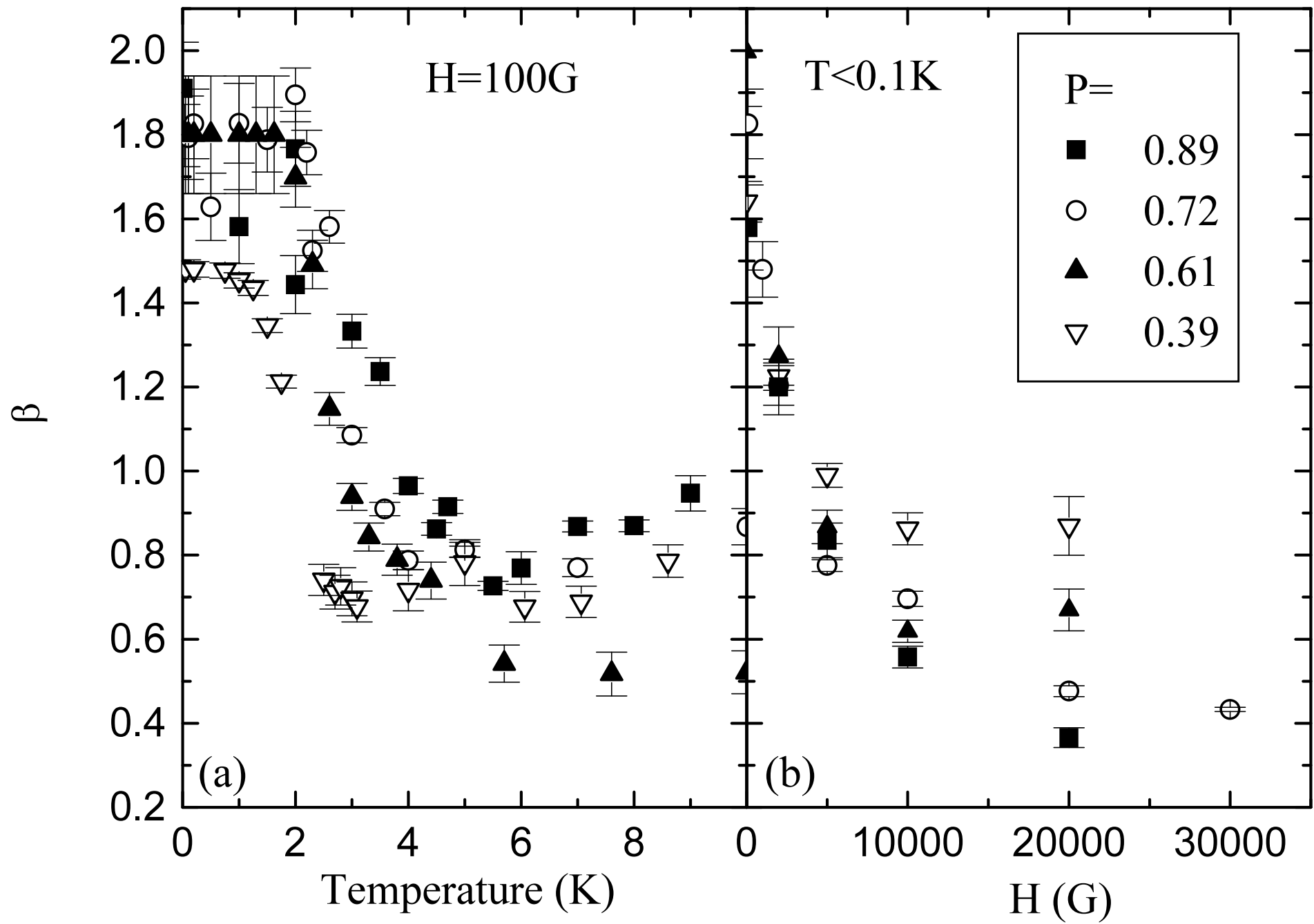
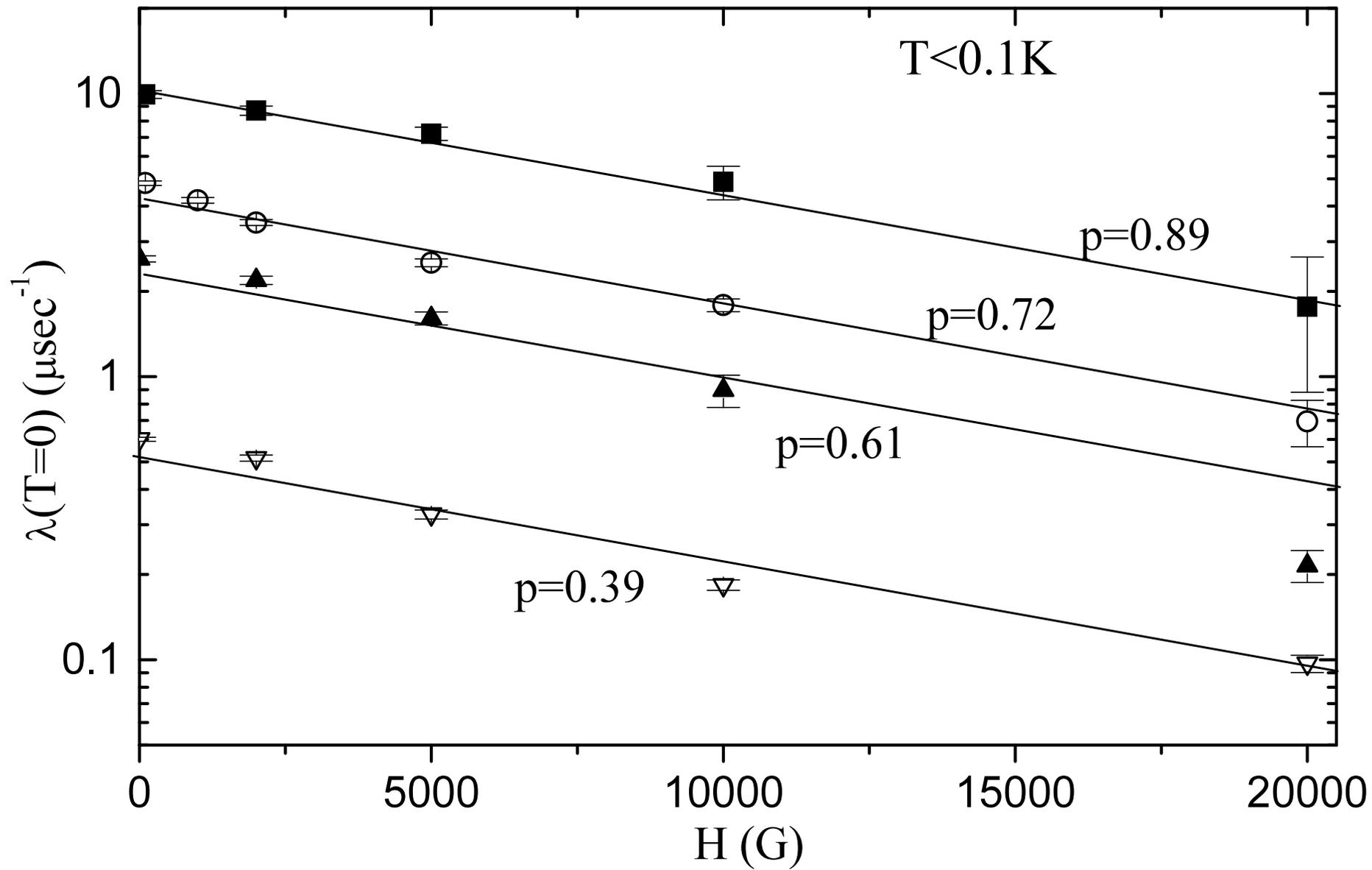
\end{document}